\input{psfig.sty}

\documentclass[12pt,preprint]{aastex}

\newcommand{\be}{\begin{equation}}
\newcommand{\ee}{\end{equation}}
\newcommand{\bea}{\begin{eqnarray}}
\newcommand{\eea}{\end{eqnarray}}

\shorttitle{A theoretical study of the luminosity temperature
relation} \shortauthors{A. Del Popolo et al.
}

\begin{document}


\title{A theoretical study of the luminosity temperature relation for clusters of galaxies}


\author{A. Del Popolo\altaffilmark{1}, N.Hiotelis\altaffilmark{2} \& J. Pe\~{n}arrubia\altaffilmark{3}
}

\altaffiltext{1}{Bo$\breve{g}azi$\c{c}i University, Physics
Department,
     80815 Bebek, Istanbul, Turkey}

\altaffiltext{2}{ First Experimental Lyceum of Athens, Ipitou 15,
Plaka 10557, Athens, Greece} 

\altaffiltext{3}{Astronomisches Rechen-Institut, Mönchhofstrasse 12-14, D-69120 Heidelberg, Germany
}


\begin{abstract}

A luminosity-temperature relation
for clusters of galaxies
is derived. The two models used, take into account the angular
momentum acquisition by the proto-structures during their expansion
and collapse. The first one is a modification of the self-similar
model (SSM) while the second one is a modification of the
Punctuated Equilibria Model (Cavaliere et al. 1999). In both
models the mass-temperature relation (M-T) used is based on the
calculations of Del Popolo (2002b).\\
We show that 
the above models lead, in X-rays, to a luminosity-temperature
relation that scales as $L \propto T^5$, at scale of groups,
flattening to $L \propto T^3$ for rich clusters and converging to
$L \propto T^2$ at higher temperatures. However a fundamental
result of our paper is that the non-similarity in the L-T
relation, can be explained by a simple model that takes into
account the amount of the angular momentum  of a proto-structure.
This result is in disagreement with the widely accepted idea that
the above non-similarity is due to non-gravitating processes as
those of heating/cooling.

\end{abstract}



\section{Introduction}

Observations of clusters of galaxies (e.g, ROSAT, ASCA), performed in the past decade,
have shown the existence of a tight correlation between the total gravitating mass of
clusters, $M_{\rm tot}$, their X-ray luminosity ($L_{\rm X}$),
and temperature ($T_{\rm X}$) of the intra-cluster medium (ICM) (David et al. 1993; Markevitch 1998; Horner,
Mushotzky \& Sharf 1999 (hereafter HMS)).
The importance of these relations is due to the fact that cluster masses are difficult to measure directly, and when comparing cluster observations with models of structure formation a surrogate for cluster mass is used. Since $M_{\rm tot}$ compares with the ICM temperature measurements which can be obtained through X-ray spectroscopy, this explains the importance of an M-T relation.
On the one hand, the X-ray temperature measures the depth of the potential wells, and
the bolometric luminosity, $L \propto n^2 R^3_{\rm X} T^{1/2}$, emitted as thermal bremsstrahlung by intra-cluster plasma measures the baryon number density, $n$ within the volume $R^3_{\rm X}$. Till some years ago, the cluster structure was considered to be
scale-free, which means that the global properties of clusters, such as halo mass, luminosity-temperature, and X-ray luminosity would scale self-similarly (Kaiser 1986). In particular, the gas temperature would scale with cluster mass as $T \propto M^{2/3}$ and the bolometric X-ray luminosity would scale with temperature as $L \propto T^{2}$, in the bremsstrahlung-dominated regime above $\sim 2$ keV.
\footnote{Indeed,
numerical simulations that include gas dynamics but exclude
non-gravitational processes such as radiative cooling and supernova
heating produce clusters that obey these scaling laws (e.g., Evrard,
Metzler, \& Navarro 1996; Bryan \& Norman 1998; Thomas et. al 2001b).}

Studies following that of Kaiser (1986) showed that
the observed luminosity-temperature relation is closer to $L \propto T^3$
(e.g., Edge \& Stewart 1991), indicating that non-gravitational processes
should influence the density structure of a cluster's core, where most
of the luminosity is generated (Kaiser 1991; Evrard \& Henry 1991;
Navarro, Frenk \& White 1995; Bryan \& Norman 1998).
One way to obtain a scaling law closer to the observational one is to have non-gravitational energy injected into intra-cluster
medium (ICM) before or during cluster formation. This solution, called pre-heating, was originally invoked to
solve two related problems: a) to explain (Kaiser 1991; Evrard \& Henry 1991) \footnote{Kaiser's self similar model predicts $L \propto T^{3.5}$. Evrard \& Henry (1991) obtained the relation $L \propto T^{11/4}$} the apparent negative evolution of the X-ray cluster luminosity function (Gioia et al. 1990; Henry et al. 1992) from the Einstein Medium Sensitivity Survey in a $\Omega_{\rm m}=1$ Universe ; b) to explain (White 1991) \footnote{In this case pre-heating was in form of supernovae-driven galactic winds} why groups and low-mass clusters seem to have higher X-ray temperatures than expected based on member velocity dispersions.

{\bf The mechanisms proposed to explain the slope change of the L-T relation can be divided into three main categories: \\
(i) models that include a pre-heating of the gas within a cluster.}
Ponman et al. (1999) showed that the entropy of the ICM in the centre of low-temperature clusters is greater than the value expected from gravitational collapse.
It has been shown that models that include an additional gas entropy can successfully reproduce many observational properties (Bower et al. 1997; Cavaliere et al. 1997, 1999; Tozzi \& Norman 2001; Borgani et al. 2001; Voit \& Brian 2001).\\
{\bf(ii) models that implement feed-back processes that alter the gas characteristics during the evolution of the cluster.}
In principle, there are many different physical processes that could break the self-similar scaling, including heating from SN or from AGN, or the removal of low-entropy gas via radiative cooling with subsequent supernova heating (Voit \& Bryan 2001).
Another example is that of Muanwong et al. (2001), who simulated galaxy cluster formation including radiative cooling with cool gas dropout and was able to reproduce the $L \propto T^3$ dependence, without adding any entropy to the gas. Moreover
other possibilities, such as magnetic pressure or cosmic-ray pressure have not been ruled out.
Allen \& Fabian (1998) have examined the effects of cooling flows for a sample of the most X-ray luminous ($L_{\rm Bol}>10^{45}$ erg/s) finding a flattening from $L \propto T^3$ to $L \propto T^2$ in agreement with models that include the effects of shocks and pre-heating on the X-ray gas (Cavaliere et al. 1997, 1999).
Cavaliere et al. (1997, 1998), have constructed a model in which the observed L-M relation on both cluster and group scales can be reproduced by varying the gas density at the virial radius, according to the accretion-shock strength, as determined by the temperature difference between the infalling and virialised gases.
Another possibility to explain the L-T relation are systematic variations in the baryonic fraction with cluster mass (David et al. 1993).
To distinguish among these processes, 
{\bf observations of high-redshift groups and clusters will be crucial to measure the evolution of the observed scaling relations as function of redshift }.\\
{\bf (iii) hydro-dynamical models that do not include gas pre-heating, nor feed-back processes, which also reproduce the available observational data (e.g., Bryan \& Norman 1998). Throughout this paper, we will analyse this last scenario.} 

On the other hand, the mass-temperature relation, seemed like it ought to be more fundamental
and less sensitive to non-gravitational effects. Yet, observations collected over the last few years
indicate that this relation also disagrees with both the scale-free predictions
and simulations that exclude non-gravitational processes (Horner,
Mushotzky, \& Scharf 1999; Nevalainen, Markevitch, \& Forman 2000 (hereafter NMF);
Finoguenov, Reiprich, \& B\"ohringer 2001 (hereafter FRB); Xu, Jin, \& Wu 2001).  These
results derive mostly from resolved X-ray and temperature
profiles coupled with the assumption of hydrostatic equilibrium,
{\bf and} 
they do seem consistent with gravitational lensing measurements (Allen, Schmidt, \& Fabian 2001).
Understanding the scaling properties of clusters is of broad
importance because these scaling laws are integral to
determination of cosmological parameters.
Thus, any inaccuracies
in the mass-temperature relation propagate into uncertainties in
cosmological parameters derived from clusters {\bf (e.g., Voit 2000, hereafter V2000)}.


In Del Popolo (2002b), we derived the mass-temperature relation and its time evolution for clusters of galaxies
in different cosmologies. {\bf We use} two different models: the first one is a modification and improvement of a model by Del Popolo \& Gambera (1999), based upon a modification of the top-hat model in order to 
account {\bf for} angular momentum acquisition by proto-structures and for an external pressure term in the virial theorem.
The second one is an improvement of a model proposed by V2000, {\bf again to account for the angular momentum acquired by proto-structures during their formation.}
Both models showed that the M-T relation is not self-similar. A  break is present in the quoted relation at $T \sim 3 {\rm keV}$ and, at the lower mass end, the power law index of the M-T relation is larger than $\alpha=3/2$ even in flat universes.
The slope of the power-law index depends on the considered cosmology.
The two models also agree in predicting a more modest time evolution of the quoted relation in comparison with the results of previous models, {\bf which also depends on the cosmology}.\\
This is in agreement with studies showing that the self-similarity in the M-T relation seems to break at
some keV (NMF; Xu, Jin \& Wu 2001). By means of ASCA data, using a small sample of 9 clusters (6 at 4 keV and 3 at $\sim 1$ keV), NMF has shown that $M_{\rm tot} \propto T_{\rm X}^{1.79 \pm 0.14}$ for the whole sample, and
$M_{\rm tot} \propto T_{\rm X}^{3/2}$ excluding the low-temperature clusters. Xu, Jin \& Wu (2001) has found $M_{\rm tot} \propto T_{\rm X}^{1.60 \pm 0.04}$ (using the $\beta$ model), and $M_{\rm tot} \propto T_{\rm X}^{1.81 \pm 0.14}$ by means of the Navarro, Frenk \& White (1995) profile. FRB have investigated the T-M relation in the low-mass end finding that $M \propto T^{\sim 2}$, and $M \propto T^{\sim 3/2}$ at the high mass end. This behaviour has been attributed to the effect of the formation redshift (FRB) (but see Mathiesen 2001 for a different point of view), or to cooling processes (Muanwong et al. 2001) and heating (Bialek, Evrard \& Mohr 2000).
Afshordi \& Cen (2001) (hereafter AC) have shown that non-sphericity introduces an asymmetric, mass dependent, scatter for the M-T relation altering its slope at the low mass end ($T \sim 3$ keV).\\
The L-T and M-T relations are somehow related: as shown by Shimizu et al. (2003),
it is possible 
to make a reliable prediction for the $L$-$T$
relation once the $M$-$T$ relation is specified. In turn, one can
obtain the $M$-$T$ relation that reproduces the observed $L$-$T$
relation without assuming an ad hoc model for the thermal evolution of
intra-cluster gas.
The two relations (M-T and L-T) are strictly connected, as shown by Shimizu et. al (2003).

%
%
This conclusion
in turn indicates that the $L$-$T$ relation provides a good
diagnosis of the underlying $M$-$T$ relation, which is as yet poorly
determined observationally.

Apart from the physical mechanism of the
additional thermal processes, there are three {\bf effects} that might
modify the mass dependence of X-ray luminosity
and steepen the resulting $L$-$T$ relation. First, the gas density
profile might be significantly flatter for less massive systems.
Second, the mass dependence of the hot gas mass fraction is strong as
$f_{\rm gas} \propto M_{\rm vir}^{1/3}$. Finally, the mass-temperature
relation is $T_{\rm gas} \propto M_{\rm vir}^{2/5}$. In practice, a
realistic model should be a combination of those three effects to some
extent.
%
%

In this paper we derive a luminosity-temperature relation for
clusters of galaxies  that takes into account the amount of the
angular momentum of proto-structures. We use two different models: the first (that we call MSSM) is a
modification of the self-similar model (SSM) while the second one
is a modification of the Punctuated Equilibria Model (hereafter
MPEM) (Cavaliere et al. 1999). We show that the presence of the
angular momentum during the gravitational collapse leads to
non-self similar L-T relation.
The two models used are described in  Sect. 2. The results are
presented and discussed in Sect. 3 and the conclusions are
summarised in Sect.4.
\section{Model}

\subsection{Modified Self-Similar model for the L-T relation}

The L-T relation constitutes a fundamental link between the
physics of the baryon component and the dynamical properties of
the Dark Matter condensations. The simplest model describing {\bf that} 
relation is the SSM model (Kaiser 1986), obtained assuming
that the gas density or the baryon number density, $n$, is
proportional to the average Dark Matter density, $\rho$, and that
the virial radius, $R_{\rm vir}$ is proportional to $R_{\rm X}$
(see introduction for a definition). In this way, one obtains,
according to this last, $L \propto M_{\rm vir} \rho T^{1/2}$. In
fact, $T \propto M_{\rm vir}/R_{\rm vir}$, $n \propto \rho \propto
M_{\rm vir}/R_{\rm vir}^3$, $R_{\rm vir} \propto R_{\rm X}$ and $L
\propto \int_0^{R_{\rm vir}} \rho^2 T^{1/2}r^2 dr \propto \rho^2
T^{1/2} R_{\rm vir}^3$, and recalling that $R_{\rm vir} \propto
(M_{\rm vir}/\rho)^{1/3}$, leads to $L \propto \rho M_{\rm vir}
T^{1/2}$ or recalling that $R_{\rm vir} \propto (T/\rho)^{1/3}$,
we get $L \propto \rho^{1/2} T^2$. This last result is
inconsistent with observed correlation close to $L \propto T^3$
(Edge \& Stewart 1991; Mushotzky 1994). Additionally, a further
steepening at the temperature of galaxy groups is indicated for
thermal  emission not associated with single galaxies (Ponman et
al. 1996).

In the following, we derive
a modified SSM, showing
that slope of the L-T relation changes at different scales.

Using Balogh et al (1999) notation, let us begin with a cluster with gas temperature $T(r)$, density profile $\rho(r)_{\rm g}$,
for which the bolometric X-ray luminosity from Bremsstrahlung scales as:
\begin{equation}
L ={6 \pi k \over C_1(\mu m_{\rm p})^2}
\int_0^{R_{\rm vir}}r^2\rho_g(r)^2T_g(r)^{1/2}dr
\label{eq:bol}
\end{equation}
%
%
(see Balogh et al. 1999),
where $C_1=3.88 \times 10^{11}$s K$^{-1/2}$ cm$^{-3}$,
$\mu=0.59$,
$R_{\rm vir} \propto (M_{\rm vir}/\rho)^{1/3}$ is the virial radius
where $\rho(z) \propto (1+z)^3$ is the Dark Matter (DM) density in the cluster, proportional to the average cosmic DM density $\rho_{\rm u}$ at formation. The simplest model describing the L-T relation, that can be calculated by
Eq. (\ref{eq:bol}), is the SSM (Kaiser 1986), assuming that the $\rho_{\rm g} \propto \rho$.

We only consider haloes in which
not all of the gas within $R_{\rm vir}$ has had time to cool since the halo formed.

We assume a singular, truncated isothermal sphere for the dark
matter potential, $\rho(r)=\rho_R(r/R_{\rm vir})^{-2}$, where
$\rho_R$ is the density at the virial radius, $R_{\rm vir}$, and
is equal to a third of the mean density {\em within} $R_{\rm
vir}$, $\bar{\rho}(R_{\rm vir})$. This latter quantity is related
to the critical density at redshift z by $\bar{\rho}(R_{\rm
vir})=\Delta_c(z)\rho_c(z)$, and $\Delta_c = 78 \Omega(z) + 80 +
300\Omega(z) / (1+15\Omega(z))$ is a fit, accurate to better than
2 per cent, to the results of the spherical collapse model as
presented in Eke et al. (1996).  It will be convenient to define a
redshift evolution term, $F_1(z)^2=(1+z)^2(1+\Omega_\circ
z)\Delta_c(z)/\Delta_c(0)$, so that
\begin{equation}\label{eqn-rhovir}
\rho_R={1 \over 3}\Delta_c(0)\rho_c(0)F_1(z)^2.
\end{equation}
For $\Omega_\circ=1$, $F_1(z)^2=(1+z)^3$ and $\Delta_c=178$.
In this model, we make the common assumption (e.g., Eke et al. 1996) that the gas is distributed isothermally,
with a temperature equal to the virial temperature of the halo.
If the gas is dissipationless,
its density profile will match that of the dark matter, i.e.,
\begin{equation}
\rho_g(r)=\rho_{\rm g,R}(r/R_{\rm vir})^{-2},
\label{eq:iso}
\end{equation}
and $\rho_{\rm g,R}/\rho_R=\Omega_b/\Omega_\circ$. To avoid the
singularity at $r=0$ when integrating over the assumed isothermal
profile, an arbitrary core radius of $r_c = f_c R_{\rm vir}$ is
adopted with $f_c=0.1$, such that $\rho_g(r<r_c)=\rho_g(r_c)$. The
integral in Eq. (\ref{eq:bol})  is dominated by the contribution
from within a few core radii, and thus the scaling properties of
this integral depend weakly on the assumed density profile.
Furthermore, departures from the standard profile can be
accommodated by redefining the core radius of the system.\\

In order to obtain the luminosity--mass relation,
we evaluate Eq. (\ref{eq:bol}) and use the M-T relation, 
which takes into account the angular momentum of the
proto-structure, obtained in Appendix A:
\begin{equation}
kT \simeq 8 keV \left(\frac{M^{\frac {2}{3}}}{10^{15}h^{-1}
M_{\odot}}\right) \frac{ \left[ \frac {1}{m_1}+\left(
\frac{t_\Omega }t\right) ^{\frac {2}{3}} +\frac{K_1(m_1,x)}{M^{8/3}}
\right] } { \left[ \frac {1}{m_1}+\left( \frac{t_\Omega
}{t_{0}}\right) ^{\frac {2}{3}}
 +\frac{K_0(m_1,x)}{M_0^{8/3}}
\right]
}
\label{eq:kT1}
\end{equation}
where 
$K_1(m_1,x)$ \footnote{$K_0(m_1,x)$ indicates that 
$K_1(m_1,x)$ must be calculated
assuming $t=t_0$} 
is given by:
\begin{eqnarray}
K_1(m_1,x)&=&\left (m_1-1\right )F x {\it LerchPhi} (x,1,3m_1/5+1)-
\nonumber \\
& & \left (m_1-1\right )F{\it LerchPhi}(x,1,3m_1/5)
\end{eqnarray}
and
\begin{equation}
F=\frac{2^{7/3} \pi^{2/3} \xi \rho_{\rm b}^{2/3}}{3^{2/3} H^2 \Omega}
\int_0^r \frac{{\cal L}^2 dr}{r^3}
\end{equation}
$m_1=5/(n+3)$, $t_\Omega =\frac{\pi \Omega _0}{H_o\left( 1-\Omega
_0-\Omega _\Lambda \right) ^{\frac 32}}$,
$x=1+(\frac{t_{\Omega}}{t})^{2/3}$ which is connected to mass by
$M=M_{\rm 0} x^{-3 m_1/5}$ (V2000) and $\xi=\frac{r_{ta}}{x_1}$,
where $r_{ta}$ is the turn-around radius and $x_1$ is defined by
the relation $M=\frac{4\pi\rho_bx^3_1}{3}$ with $\rho_b$ is the
background density. 
Finally we get:
%

\begin{equation}
L={ 3.31 \times 10^{45}}\,\left ({\frac {M}{M_{{o}}}}\right )^{4/3
}\left({\frac {1}{178}}\,\Delta_{{c}}\right){F_1}^{2} \left({\frac
{{\Omega_{{b}}}}{{ \Omega}}}\right)^2 \sqrt { \frac{
\frac{1}{m_1}+\left ({\frac {t_{{\Omega}}}{t}} \right )^{2/3}+
\frac{K_1(m_1,x)}{ \left ({\frac {M}{M_{{o}}}}\right )^{8/3} } }{
\frac{1}{m_1}+\left ({\frac {t_{{\Omega}}}{t}}\right
)^{2/3}+{\frac {K_0(m_1,x)}{{M_{{0}}}^{8/3}}}} } \left({\frac
{1-{\it fc}}{{\it fc} }}\right) \label{eq:self}
\end{equation}
which differently from Kaiser's (1986) prediction is not self-similar.
It reduces to the self-similar
form ($L \propto M^{4/3}$) if angular momentum acquisition is not taken into account, namely if
${\cal L} \rightarrow 0$ (or $F \rightarrow 0$).

The previous computation depends on the value of the angular 
momentum acquired by the DM haloes from tidal torques from 
surrounding matter. This enters the L-T relation through the 
quantities $F$ and $K_1$ (see also Appendix A). 
In the limit of vanishing angular momentum the L-T and 
the M-T relations reduce to the well-known self- similar 
forms. Then, it is important to add a discussion on 
the magnitude of the angular momentum calculated as in previous papers (e.g., Del Popolo \& Gambera 1998; 
Del Popolo et al. 2001).

The angular momentum is
acquired by the cosmological torque acting on the proto-structures
due to the tidal field of the environment. The amount of angular
momentum as well as its distribution are related to the assumed
power spectrum of density perturbations. We have to note here
that the problem of the growth of angular momentum of
proto-structures from the tidal torques of the surrounding matter
has been studied extensively in the literature with both
analytical and numerical (N-body) methods (e.g. Efstathiou \&
Jones 1979; Barnes \& Efstathiou 1987; Voglis \& Hiotelis 1989;
Warren et al. 1992; Eisenstein \& Loeb 1995; Kratsov et al. 1998).
 A main result of the above studies is that the values of the
dimensionless spin parameter $\lambda\equiv {\cal L}
|E|^{1/2}/GM^{5/2}$, (Peebles 1971), follow a log-normal
distribution with a small average value 0.05. In the above
relation ${\cal L}$ is the total angular momentum of the
proto-structure, $E$ is its binding energy, $M$ its mass and $G$
the gravitational constant. The above numerical results are
confirmed by analytical studies presented by other authors as
those of Steinmetz \& Bartelmann (1995) and Catelan \& Theuns
(1996). 
To be more precise, $\lambda$ depends on the galactic morphological type, being 
as high as $\lambda \simeq 0.5$ for spirals and SO galaxies, and 
$\lambda \simeq 0.05$ for ellipticals, although the dispersion around these values is large 
(Efstathiou \& Jones 1979). In the case of structures of $10^{12}-10^{13}$ its value is
$\simeq 0.1$ and $\simeq 0.01$ for clusters \footnote{The resulting typical circular velocities of structures
is $\simeq 150 $km/s for galaxies similar to the Milky Way, $\simeq 5$ km/s for clusters and $\simeq 10$ km/s for superclusters (see Catelan \& Theuns 1996)}.

In this paper, we calculated angular momentum as Sect. 3 in Del Popolo et al. (2001), 
following Eisenstein \& Loeb (1995). With the Bardeen et al. (1986) power spectrum smoothed on
galactic scale 
for a $\nu=2$ peak, the model gives a value of
$2.5 \times 10^{74} {\rm g cm^2/s}$, in very good agreement with Catelan \& Theuns (1996), in other words 
the amount of angular momentum used in our calculations  is consistent with the values of $\lambda$
predicted by the tidal fields of the surrounding matter. Although
this amount is in general small, our results show that {\bf it} is efficient
to lead to a non-similar L-T relation. {\bf The angular momentum of dark matter haloes } has also {\bf other important consequences}. For
example small amounts of angular momentum are able to change the
density profile of dark matter haloes from the isothermal law
$\rho(r)\propto r^{-2}$ to a profile that it flattens
significantly inwards, (e.g. Hiotelis 2002). 

Moreover, several studies have shown that the influence and the 
role of shear on structure formation is of fundamental importance. 
Shear on a density perturbation can be produced by the
intrinsic asphericity of the perturbation itself (internal shear)
or it can be due to the interaction of the perturbation with the
neighbouring ones (external shear).
For example, according to the previrialization conjecture (Peebles \& Groth 1976,
Davis \& Peebles 1977, Peebles 1990), initial asphericities and tidal interactions between neighbouring
density fluctuations induce significant non-radial motions which oppose the
collapse. This means that virialized clumps form later, with respect to the
predictions of the linear perturbation theory or the spherical collapse model,
and that the initial density contrast, needed to obtain a given final
density contrast, must be larger than that for an isolated spherical
fluctuation.
This kind of conclusion was supported by Barrow \& Silk (1981), Szalay \&
Silk (1983), Villumsen \& Davis (1986), Bond \& Myers (1993a,b)
and Lokas et al. (1996). Arguments based on a numerical least-action method lead Peebles (1990)
to the conclusion that irregularities in the mass distribution,
together with external tides, induce non-radial motions
that slow down the collapse.
In a more recent paper, Audit et al. (1997) 
they conclude that spherical collapse is the fastest.
This result is in agreement with Peebles (1990), and more recent papers, namely  
Del Popolo et al. (2001), Del Popolo (2002a). 

\subsection{Improvements to the punctuated equilibria model}

In this subsection, we shall extend the punctuated equilibria model (PEM) by Cavaliere, Menci e Tozzi (1997,1998, 1999) (hereafter CMT97, 98, 99) to take account of angular momentum acquisition from the proto-structure.
In their model (CMT98), the cluster evolution is described
as a sequence of ``punctuated equilibria" (PE),
that is to say, a
sequence of hierarchical merging episodes of the DM halos, associated in the
ICP to shocks of various strengths (depending on the mass
ratio of the merging clumps), which provide the boundary
conditions for the ICP to readjust to a new hydrostatic
equilibrium.

The X-ray bolometric luminosity of a cluster is given by Eq. (\ref{eq:bol}), which in
CMT98 notation is:
\begin{equation}
L\propto \int_o^{r_2}\,n^2(r)\,T^{1/2}(r)\,d^3r~.
\end{equation}
Here $T(r)$ is temperature in the plasma and $r_2$
is the cluster boundary, that we take to be close to
the virial radius $R_{\rm vir} \propto M_{\rm vir}^{1/3}\,\rho^{-1/3}$, where
$\rho (z)\propto (1+z)^3$ is the DM density in the cluster,
proportional to the average cosmic DM density $\rho_u(z)$ at formation.

As shown in Appendix B, the L-T relation can be casted in the form:
\begin{equation}
L\propto \Big({n_2\over n_1}\Big)^2\,
\rho\,
\Bigg[{T_2\over T_v}\Bigg]^{1/2}\,
\overline{[n(r)/n_2\big]^{2+(\gamma-1)/2}}
m^{4/3}
\sqrt {
\frac{
\frac{1}{m_{1}}+\left ({\frac {t_{{\Omega}}}{t}}
\right )^{2/3}+
\frac{K_1}{
\left ({\frac {m}{m_{{o}}}}\right )^{8/3}
}
}{
\frac{1}{m_1}+\left ({\frac {t_{{\Omega}}}{t}}\right )^{2/3}+{\frac
{K_{{0}}}{{m_{{0}}}^{8/3}}}}
}
\label{eq:lll}
\end{equation}
See Appendix B for a derivation of
Eq. (\ref{eq:lll}) and a definition
of the terms involved.

Our final aim is to compute the average value of $L$ and its dispersion,
associated with a given cluster mass $m$.

In order to reach this goal,
we must sum over the shocks produced at a time $t'<t$
in all possible progenitors $m'$ (weighting with their number)
by the accreted clumps $\Delta m$ (weighting with their merging rate);
finally, we integrate over times $t'$ from an effective lower limit
$t-\Delta t$.

The average $L$ is then given by
\begin{equation}
\langle L\rangle = Q\,\int_{t-\Delta t}^t\,dt'
\,\int_0^m\,dm'\,\int_0^{m-m'}\,d\Delta m
\,{df\over dm'}(m',t'|m,t)\,
{d^2 p(m'\rightarrow m'+\Delta m)\over d\Delta m\,dt'}
\,L~;
\label{eq:lum}
\end{equation}
and the variance is given by
\begin{equation}
\langle \Delta L^2\rangle = Q\,\int_{t-\Delta t}^t\,dt'
\int_0^m\,dm'\,\int_0^{m-m'}\,d\Delta m\,
\,{df\over dm'}(m',t'|m,t)\,
{d^2 p(m'\rightarrow m'+\Delta m)\over d\Delta m\,dt'}
\Big(L-\langle L\rangle\Big)^2~.
\label{eq:lumm}
\end{equation}
where $Q$ is the normalisation factor (the compounded probability distribution in Eqs. (\ref{eq:lum})
and (\ref{eq:lumm}) has been
normalised to 1. The effective lower limit for the integration over masses is set
as described in Sect. (2.4) of CMT99.


\section{Results}
The results of our calculation are plotted in Fig. 1-3.

\begin{figure}
\centerline{\hbox{
\psfig{file=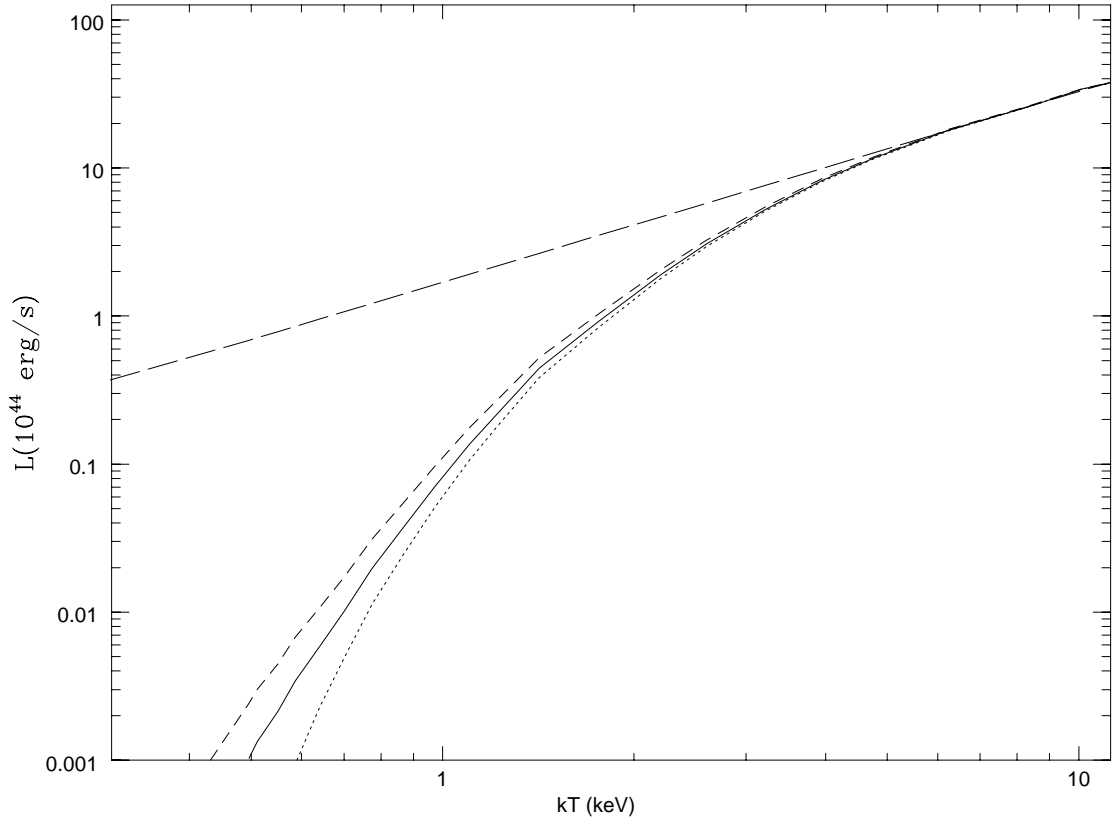,width=12cm,angle=-90}
}}
\caption[]{Comparison between the SSM, long-dashed line, with the MSSM, short dashed line, the PEM solid line,
and MPEM dotted line.}
\end{figure}

\begin{figure}
\centerline{\hbox{
\psfig{file=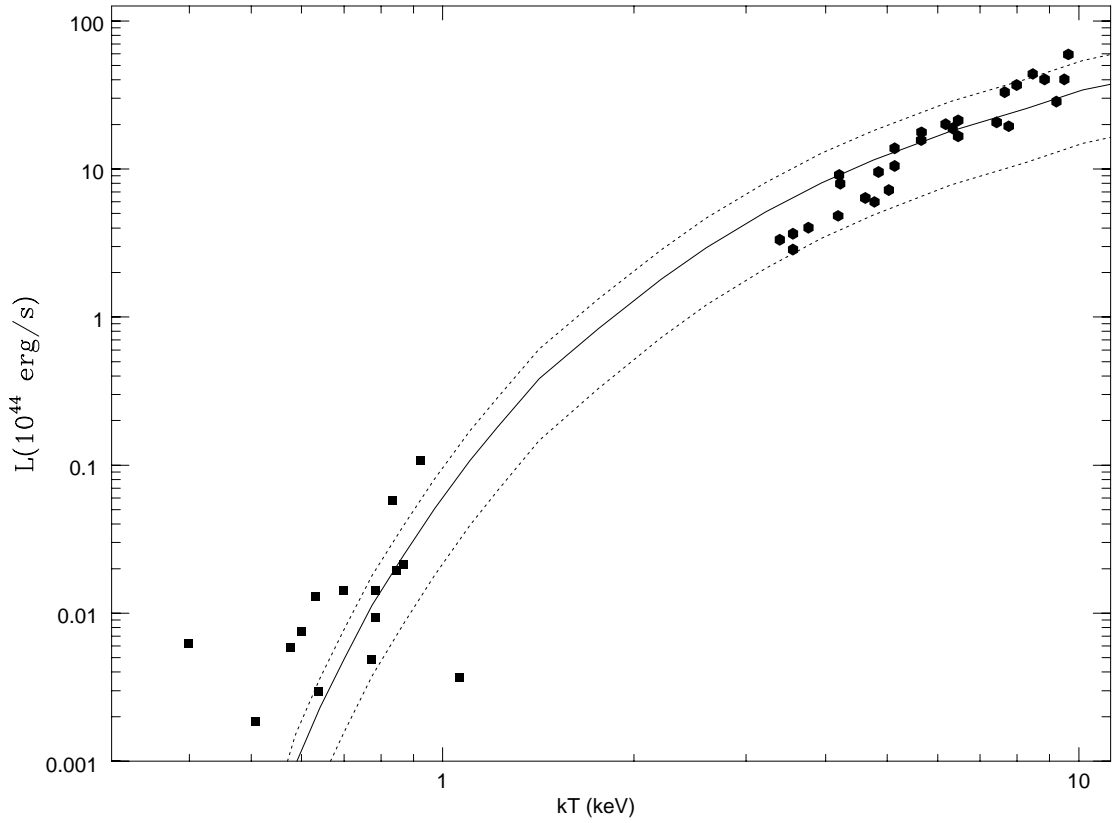,width=12cm,angle=-90}
}}
\caption[]{MPEM model: the average L-T correlation with the 2 $\sigma$ dispersion (dotted lines), for a tilted cosmogony. Group data from Ponman et al. (1996) are represented by solid squares while cluster data from Markevitch (1998) are represented by solid hexagons. }
\end{figure}

\begin{figure}
\centerline{\hbox{(3)
\psfig{file=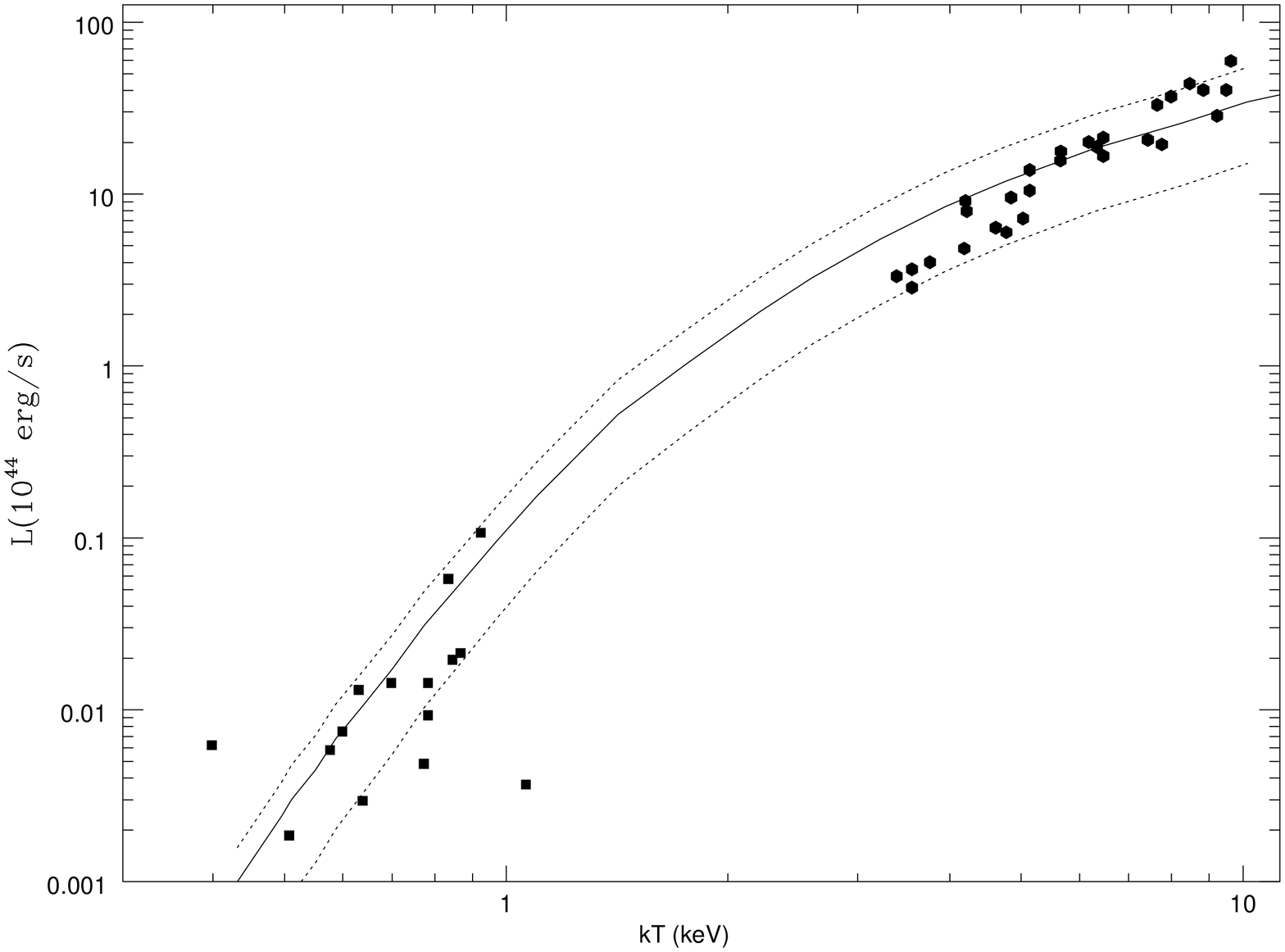,width=12cm,angle=-90}
}}
\caption[]{Similar to Fig. 2 but for the MSSM model.}
\end{figure}

In Fig. 1, we plot a direct comparison between the SSM, long-dashed line, with the MSSM,
short dashed line, with the PEM \footnote{The PEM is based on hierarchical clustering. Group and cluster formation is envisaged in terms of DM potential wells evolving hierarchically, and engulfing outer baryons by accretion of smooth gas or by merging with other clumps. After a merging episode, the ICP in the wells falls back to a new, approximate hydrostatic equilibrium. This sequence of hydrostatic equilibria of the ICP is physically motivated for all merging events except for those involving comparable clumps (a mass ratio larger than $\sim 1/4$). However these
sum up to less than  $10 \%$ in the number.
In the PEM, thermal energy of the infalling gas is initially due to stellar pre-heating (of nuclear origin); then it is increased to the virial value (of gravitational origin) when the accreted gas is bound in DM sub-clumps. So the pre-heating sets an effective threshold $kT_1 \sim 0.5$ keV to gas inclusion, which breaks the self-similar correlation $L\propto T^2$
not only in its vicinity but also up to a few keV.
}, solid line, and finally with the MPEM with a tilted CDM cosmogony.
As well known, the SSM predicts that $L \propto T^2$ (Kaiser 1986), while the MSSM predicts non self-similar behaviour of the L-T relation: namely a L-T relation $L \propto T^5$ at scale of groups, $L \propto T^3$ for rich clusters in agreement with observations and the L-T relation saturates toward $L \propto T^2$ for higher temperatures.
The plot shows that the MSSM predicts a similar behaviour of the L-T relation to that predicted by the PEM. Differences of maximum 10\% are noted for smaller values of the temperature.

It is noticed above, that a self-similar evolution for all
clusters , at typical cluster temperatures ($T>2$ keV), should
lead to  $L \propto T^2$, since free-free emission dominates the
cooling. Instead the observed L-T relation is more steep, $L
\propto T^{2.6-2.9}$, meaning that lower temperature clusters and
groups of galaxies are far less luminous than expected.
 Several different models have been proposed in order to explain the quoted behaviour in the L-T relation.
The key-point of these models is that the X-ray luminosities of low-temperature clusters
 are small because their gas is less centrally concentrated than in hotter clusters, an effect that has been attributed to an universal minimum entropy level in intra-cluster gas, resulting from supernova heating (Ponman et al. 1999; Wu et al. 2000; CMT99), from heating by active nuclei (Wu et al. 2000) or from radiative cooling (Wu et al. 2000; Bryan 2001; Pearce et al. 2000). In other terms for some reasons the core gas is less high than that expected in the self-similar model. For example, an early episode of uniformly distributed supernova feedback, could rectify the problem
  by heating the un-condensed gas and therefore making it harder to compress in the core.
In other words,
the models with pre-heating and similar gives rise to the quoted break
because they change the density in the core.
During the hierarchical buildup an energy input pre heats the gas before it falls into new groups and clusters, so hindering its flow into the latter. The core density shall decrease and so the luminosity.

A similar mechanism acts in the model of this paper.
In fact, as shown in Del Popolo \& Gambera (1998), the angular momentum acquired by a
 shell centred on a peak
in the CDM density distribution is anti-correlated with density:
 high-density peaks acquire less angular momentum
than low-density peaks (Hoffman 1986; Ryden 1988). A greater amount of angular momentum
 acquired by low-density peaks (with respect to the high-density ones) implies that
  these peaks can more easily resist gravitational collapse and consequently it is more difficult for them to
form structure, and in some conditions the structure formation by
low mass peaks is even inhibited\footnote{One interesting point to mention, at this point,
is that several different assumptions are able to reproduce the
observed L-T. This could mean that L-T relation, is not a very
sensitive test: since almost any change to the ``pure"
self-similar model, reproduces this relation.}. 
{\bf The break of the self-similarity of the L-T relation may have also important consequences for determining the cluster masses from their luminosity}.
As shown by Shimizu et al. (2003), the predicted L-T relation is very sensitive to the assumed M-T relation, and then the non self-similarity of the L-T relation is strictly connected to that in the M-T relation.
Also the M-T relation, as previously discussed, is non self-similar, and this behaviour
has been interpreted in different ways (see introduction).\\
In Del Popolo 2002b, the bent in the M-T relation is entirely justified in terms of cluster
tidal interaction with the neighbouring ones, or in other terms it is strictly connected to the asphericity of clusters (see Del Popolo \& Gambera 1999 for a discussion on the relation between angular momentum acquisition, asphericity and structure formation).
Non-sphericity introduces an asymmetric bent, dependent on mass, in the M-T relation that gives rise to a different slope at the low mass end ($T \sim 3 {\rm keV}$): the lower the mass the larger the bent.

The previous result is in agreement with AC result.
In that paper, the authors used a nearly spherical collapsing region
to obtain the M-T relation. According to their results, non-sphericity introduces an asymmetric, mass-dependent scatter
(the lower the mass, the larger the scatter) for the M-T relation, thus altering the slope at the low masses end ($T \simeq 3$ keV).

{\bf As commented in the introduction, heating/cooling mechanisms are not necessary to explain the observational L-T correlation.}
{\bf Bryan \& Norman (1998), carrying out large hydrodynamical simulations that followed the hierarchical evolution of clusters of galaxies, found that the observed M-T, L-T relations can be thoroughly reproduced if the number of particles and the spatial resolution are large enough. The importance of the numerical accuracy proves to be crucial to determining those relations, and so, for example, Mwanwong et al. (2001) using simulations with only $160^3$ particles and a spatial resolution of $100 h^{-1}$ Mpc, obtain that $L\propto T^2$ independent of mass, if radiative cooling is not implemented. In contrast, Bryan \& Norman (1998), using simulations with $512^3$ particles and a spatial resolution of $50  h^{-1}$ Mpc, did reproduce the observational bend of the L-T relation at the low-mass region.}

  {\bf {\it A priori}, it is unclear why the L-T and M-T relations are that sensitive to resolution. One possible explanation goes in the direction of the results shown in this paper. Taking into account that proto-structures gain angular momentum owing to tidal interactions with other non-spherical structures, low resolution may hinder the gain of angular momentum by preventing an accurate determination of the proto-structure shape. To clarify this point, we shall use the simulations of Mwanwong et al. (2001) and Bryan \& Norman (1998) as an example. The particle mass, $m_p$, of the first was $m_p=2.1\times 10^{10} h^{-1} M_\odot$, whereas the later used $m_p\simeq 6\times 10^{8} h^{-1} M_\odot$ in their simulations. Since a common characteristic of hydro-dynamical calculations is that all particles have the same mass, clusters with $ kT < 2$ keV (where the bend in the L-T relation starts to departure clearly from self-similarity) have masses of $M\sim 3\times 10^{14} h^{-1} M_\odot$, i.e, they contain approximately  $N<1.4 \times 10^4$ and $N<5 \times 10^5$ particles in Mwanwong et al. (2001) and Bryan \& Norman (1998) simulations, respectively. As we go to lower masses, we reduce the number of particles enclosed in proto-structures. As a consequence, the shape and, therefore, the inertia axes, may fluctuate randomly, which w
ould lead in average to a systematical decrease of angular momentum gained by low-mass proto-clusters. }

{\bf Besides the poorly determined shape of low-mass structures, one must also take into account the possible effects of force resolution.}
In a typical N-Body evolution code, like for example the Treecode 
of L. Hernquist (1987), the force acting on a particle is given
by the sum of two components: the force coming from the nearest 
neighbours and that coming from an expansion of the gravitational 
potential of the entire system up to quadrupole terms. As can be shown, 
the value of the average stochastic force in the simulation, $ F_{sim}$, 
is an order of magnitude bigger than that obtained from the theory, 
$ F_{th}$, of stochastic forces. As a consequence only the higher force are taken into account, 
while the small fluctuations induced by the small scale substructure are not 
"seen". This is the case for CDM models in which the 
stochastic force generators are substructures at least three orders 
of magnitude smaller in size than the proto-structures in which they are embedded 
(e.g. clusters of galaxies). 

{\bf Taking into account the large simulations required to obtain a good description of the L-T relation,} it is not surprising that  similar disagreements are
reported in other cases too. For example, in the case of the M-T
relation, it is noted that the results from different
observational methods of mass measurements are not consistent with
one another and with the simulation results (e.g., Horner,
Mushotzky, \& Scharf 1999, hereafter HMS; Neumann, \& Arnaud 1999;
Nevalainen, Markevitch, \& Forman 2000, FRB). In general, X-ray
mass estimates are about $80\%$ lower than the predictions of
hydro-simulations. On the other hand, X-ray mass estimates lead to
normalisations about $50\%$ higher than our result and
simulations.

One possible source for difference between theoretical and observational normalisations is that the
values for $\tilde{\beta}$\footnote{$\tilde{\beta} = \beta[1+f(1/\beta-1)\Omega_{\rm b}/\Omega_{\rm m}]$,
where $f$ is the fraction of the baryonic matter in the hot gas, and $\Omega_{\rm b}$ is the density parameter of the
baryonic matter.}
are different in the two cases due to
systematic selection effects. Also, intriguingly, Bryan \& Norman (1998) showed that there is
a systematic increase in the obtained value of $\tilde{\beta}$ by increasing the
resolution of the simulations.

We would like to stress that even {\bf if} the effects of angular momentum
are not taken into account, this last process gives rise to
self-similar structures only in a first approximation. In fact:
(a) the effective spectral index $n_{\rm eff}$ of CDM models
depends, even if weakly, from the scale, going from values of
$n_{\rm eff} \simeq -1.2$ for clusters to $n_{\rm eff} \simeq -2$
for galaxies; (b) we live in a universe with cosmological constant
different from zero, which means that there is a typical redshift
at which it became important for cosmic dynamics; (c) even dark
matter profiles are not perfectly self-similar, since they depends
on the concentration parameter, which in turn is inversely
proportional with mass, this because smaller structures formed, on
average, at earlier times, when cosmic density was larger.

As reported, Fig. 1 shows a slight difference between the SMMS prediction and that of PEM, being the slope
predicted by SMSS, at low temperatures less steep than that of PEM, and MPEM. The difference is not so large,
implying a difference in luminosity of 10\% (larger for SMSS with respect to PEM).
Also in Fig. 1, it is plotted the L-T relation predicted by MPEM.
In this last case, the bending is produced by two effects: the threshold effect of the pre-heating temperature 
$k T_1 \simeq 0.5$ keV (as in CMT99) and to the effect of angular momentum acquired by clusters. As a consequence, if we compare MPEM with PEM (or MSSM), the bending is larger (beside the threshold effect we have the acquisition
of angular momentum).

Relative to this last item, looking at Fig. 1, one can see that the curve obtained from MSSM is very different to that corresponding to SSM, whereas the one from MPEM differs not much to that of PEM. The reason is the following: if we consider a cluster, without implementing pre-heating, the angular momentum acquisition is responsible for the slowing down and eventual stopping of the matter collapse towards the centre of the cluster, leading to the discussed consequences. Implementing pre-heating, this gives rise
(by heating the uncondensed gas and therefore making it harder to compress in the core) to a region at higher temperature and pressure that acts like a boundary for the infalling gas which, therefore, reduces the effects induced by angular momentum acquisition.


In Fig. 2, we plot the results for the MPEM model: the average L-T correlation with the 2 $\sigma$ dispersion (dotted lines), for a tilted cosmogony. Group data from Ponman et al. (1996) are represented by solid squares while cluster data from Markevitch (1998) are represented by
solid hexagons.
The L-T correlation is given by the double convolution (Eq. (\ref{eq:lum})), while $\Delta L$ is obtained by
Eq. (\ref{eq:lumm}). The normalisation has been fitted on the data (see CMT99).

The quantities and profiles of the PEM model are the same of
CMT99, namely the reference cluster has a mass $m=M/M_0$, and a
dark matter potential $\phi(r)$ as described in Appendix B. The
density and temperature profiles are given by Eq. (\ref{eq:dens})
and they should match the shock boundary conditions at the
position $r_2 \simeq R_{\rm vir}$. The average value and scatter
of the parameter $\beta$, given by Eq. (\ref{eq:mu}), calculated
through the PEM and shown in Fig. 2 of CMT99, increases from
$\beta=0.5$ to $\beta \simeq 0.9$ while the baryonic fraction
$f_2$ is the one in Fig. 3 of CMT99. The $\gamma$ parameter is
fixed as described in Appendix B.  As the plot shows, in agreement
with CMT99, the correlation is not a simple power law, but it
starts as $L \propto T^2$ for very rich clusters, and after it
bends down with decreasing T. As previously told, the bending is
induced by two {\bf mechanisms: the threshold imposed by the pre-heating
temperature $k T_1 \simeq 0.5$ keV (as in CMT99) and the 
angular momentum acquired by clusters.}
As a consequence, if we compare MSSM with PEM, the bending is larger (beside the threshold effect we have the acquisition
of angular momentum).

We want to point a similitude between the role of pre-heating
temperature, $T$, in the PEM and {\bf that of the} angular momentum $\cal L$ in our
model. In the PEM, the thermal energy of infalling
gas comes initially from stellar pre-heating (of nuclear origin);
then it is increased to the virial value (of gravitational origin)
when the accreted gas is bound to DM sub-clumps. So the pre-heating
sets an effective threshold $kT_1 \sim 0.5$ keV to gas inclusion,
which breaks the self-similar correlation $L\propto T^2$ not only
in its vicinity but also up to a few keV. Increasing the
pre-heating temperature the bending in the L-T relation becomes
more pronounced. A similar process occurs if the acquired angular
momentum is larger.


A similar fitting formula to that of CMT98, for the predicted L-T correlation (for $T>T_1 \simeq 1$ keV), is given by:
\begin{eqnarray}
L & = & a_L\,T^{2+\alpha_L}\,(\rho/\rho_o)^{1/2} \\
a_L & \propto &
\,\Omega_0^{0.3}\,(1+z)^{0.22/\Omega_0}+(1-\Omega_0)\,e^{-0.7\,(1+z)}
\nonumber
 \\
\alpha_L & = &
a_1\,(1+z)^{-0.2}\,e^{-a_2\,(T-T_1)/\Omega_0^{0.1}\,(1+z)^{0.5}} ~,
\nonumber
\label{eq:fit}
\end{eqnarray}
where the luminosity is expressed in units of $10^{44}$ erg/s and the
temperature in keV, and with $a_1=1.2$ and $a_2=0.17$.

%

At temperatures larger than the threshold $kT_1 \simeq 0.5$ keV, the relative $\Delta L/L$ remains constant around $25 \%$.
A study of the dependence of $\langle L\rangle$ and $\Delta L$ on $\Omega_0$ shows that both these quantities increase with increasing $\Omega_0$, similarly to what shown in CMT99 \footnote{This is because the underlying strength of the current shocks grows on average as the merging rate (moderately) increases on approaching the
critical cosmology, see Lacey \& Cole (1993).}

Fig. 3, shows MSSM compared with observational data. Similarly to Fig. 2, we plot the average L-T correlation with the 2 $\sigma$ dispersion (dotted lines), for a tilted cosmogony. Group data from Ponman et al. (1996) are represented by solid squares while cluster data from Markevitch (1998) are represented by solid hexagons.
The L-T correlation can be fitted in this case by a similar formula to that of Eq. (\ref{eq:fit}), with
$a_1=1.28$ and $a_2=0.19$.
As reported, Fig. 1 shows a slight difference between the SMMS prediction and that of PEM, being the slope
predicted by SMSS, at low temperatures less steep than that of PEM. The difference is not so large, implying a difference in
luminosity of 10\% (larger for SMSS). The fit to data of SMSS model, as Fig. 3 shows, is also very good.


{\bf To summarise, the key idea of the SMSS model and of other mechanisms proposed to reproduce the non-self-similarity of the L-T relation is in all cases fairly similar:} If one wants to have clusters less luminous than SSM prediction, it is necessary to have
 a physical process that reduces the quantity of gas infalling towards the centre of the cluster which, therefore, reduces the core luminosity. {\bf In the case of heating/cooling models, some energy input pre-heats the gas before it falls into new groups and clusters, hindering its flow into the latter. In the SMSS model, that role is played by the initial spin present in proto-clusters. }


\section{Conclusions}

In this paper we showed that the  presence of angular momentum
during the collapse of a proto-structure leads to a
non-self-similar L-T relation. The quoted effect leads, in X-rays,
to a luminosity-temperature relation that scales as $L \propto
T^5$, at scale of groups, flattening to $L \propto T^3$ for rich
clusters and converging to $L \propto T^2$ at higher temperatures.

These results  are in disagreement with the largely
accepted assumption that heating/cooling processes and similar are
fundamental in the originating the non-self similar behaviour
(shaping) of the L-T relation. {\bf As Bryan \& Norman (1998) showed, it is not necessary to
hypothesise pre-heating/cooling models in order to reproduce observations, on the contrary, it is possible to reproduce the observed L-T relation if the spatial and mass resolution are accurate enough. Poorly resolved clusters, with few particles enclosed, lead to self-similar L-T curves.}

We have shown that the large bend of the L-T relation is caused by the fact that the angular momentum acquired by a
shell centred on a peak in the CDM density distribution is
anti-correlated with density: high-density peaks acquire less
angular momentum than low-density peaks. A greater amount of
angular momentum acquired by low-density peaks (with respect to
the high-density ones) implies that these peaks can more easily
resist gravitational collapse and consequently it is more
difficult for them to form structure. This results in a tendency
for less dense regions to accrete less mass with respect to a
classical spherical model. As a consequence, the X-ray luminosities of
low-temperature clusters are small because their
gas is less centrally concentrated than in hotter clusters.



\newpage

\section{Appendix A: M-T relation}

As previously quoted, numerical methods and simple scaling arguments suggest that the X-ray temperature of clusters,
$T_{\rm X}$, can be directly related to their masses
as $M_{\rm vir} \propto T_{\rm X}^{3/2} \rho_{\rm b}^{-1/2} \Delta_{\rm vir}^{-1/2}$,
where $\rho_{\rm b}$ is the critical density, $\Delta_{\rm vir}$ the mean density within the virial radius $R_{\rm vir}$.

In Del Popolo \& Gambera (1999) and Del Popolo (2002),
we got the M-T relation in two different ways: (1) modifying the top-hat model; (2) modifying Voit \& Donahue (1998) (hereafter V98) model.\\
In the first, we modified the top-hat model in order to take account of angular momentum acquisition by protostructures and used a modified version of the virial theorem in order to include a surface pressure term (V2000, AC). This correction is due to the fact that at the virial radius $R_{\rm vir}$ the density is non-zero and this requires a surface pressure term to be included in the virial theorem (Carlberg, Yee \& Ellingson 1997) (the existence of this confining pressure is usually not accounted for in the top-hat collapse model).
The derivation of the previous relation
is fundamentally based on the approximation of cluster formation with the evolution of a spherical top-hat density perturbation (Peebles 1993) and on the additional assumption that each cluster observed at a redshift $z$ has just reached the moment of virialization. This last assumption is currently known as the late-formation
approximation, which is a good one in a critical $\Omega_{\rm 0}=1$, because for this value of $\Omega$ massive clusters develop rapidly at all redshifts and the moment of virialization is always close to that of observation. In other terms for $\Omega_{\rm 0}=1$, the accretion rate remains sufficiently high, and this implies that the clusters we actually observe attained their observed masses recently. In the $\Omega_{\rm 0}<1$ case cluster formation is ``shutting down"  and it is necessary to take account of the differences between the moment of virialization and that of observation. The problem becomes worse going through $\Omega_{\rm 0}<<1$: in fact in the late-formation approximation $M_{\rm vir}$ rises steadily  since $\rho_{\rm b} \Delta_{\rm vir}$ declines indefinitely, while we expect that the cluster formation is going to stop \footnote{The result of the late-formation approximation is displayed in Eqs. 18-19 of Del Popolo (2002)}.  \\

The late-formation approximation is a good one for many purposes, but a better one can be obtained in the low-$\Omega$ limit. As can be found in the literature, there are two ways of improving the quoted model. One is to define a formation redshift $z_{\rm f}$ at which a cluster virializes and after the properties of observed clusters at $z$ are obtained by integrating over the appropriate distribution of formation redshifts (Kitayama \& Suto 1996; Viana \& Liddle 1996). The second possibility is the one described by V98, V2000. In this approach, the top-hat cluster formation model is substituted by a model of cluster formation from spherically symmetric perturbations with negative radial density gradients. The fact that clusters form gradually, and not instantaneously, is taken into account in the merging-halo formalism of Lacey \& Cole (1993). In hierarchical models for structure formation, the growth of the largest clusters is quasi-continuous since these large objects are so rare that they almost never merge with another cluster of similar size (Lacey \& Cole 1993). So, Lacey \& Cole (1993) approach extends the Press-Schechter formalism by considering how clusters grow via accretion of smaller virialized objects.
Summarising, in order to obtain the proper normalisation and time evolution of the M-T relation, one has to account:\\
a) for the continuous accretion of mass of clusters;\\
b) for the non-zero density at $R_{\rm vir}$, requiring a change in the virial theorem by including a surface pressure term.

The M-T relation derived by means of a model of continuous accretion, differs from the late-formation model in both normalisation and time-dependent behaviour. \footnote{A comparison of the normalisation predicted by the late-formation model with that predicted by simulations of Evrard, Metzler \& Navarro (1996) shows that when $\Omega_0=1$ this normalisation is only 4\% below the empirical value, but it lies 20\% below it for $\Omega_0=0.2$.
In the case of V2000 model and for a power-law spectrum, a comparison with the same simulations show that the temperature normalisation of the $n=-2$  case deviates by less than 10\% over the range $0.2<\Omega_{\rm 0}<1$ and by $\simeq 18\%$ in the case $n=-1$ (V2000). The normalisation obtained by the V2000 model, even if it is more accurate than that given by the late formation, or that by AC which is in agreement with hydro-simulations, show a noteworthy discrepancy when compared with X-ray mass estimates (about 50\% for the AC model; see also V2000).
One possible source for differences in theoretical and observational normalisations may be due to the fact that $\beta$ is different in the two cases because of systematic selection effects. For example, as shown by Bryan \& Norman (1998), increasing the resolution of simulations there is an increase in the value of $\beta$.
So summarising, for what concerns normalisation,
the continuous formation model gives more precise results than the late formation one, but in any case if we want to fit observations we need to shift the normalisation (see AC).
%
}

In order to obtain the M-T relation in this second approach,
we assume, as shown by V2000, the mass grows like $M \propto \omega^{-3/(n+3)}$ (Lacey \& Cole 1993; V98; V2000). The virial energy of the cluster, $-E$, can be calculated by integrating the 

In order to obtain an expression for the kinetic energy, we firstly calculated $E/M$:
\begin{equation}
\frac{E}{M}=-\frac{\int \epsilon dM}{M}=
\frac{3m_1}{10(m_1-1)}\left( \frac{2\pi G}{t_\Omega }\right)
^{\frac 23}M^{\frac 23} \left[ \frac 1m_1+\left( \frac{t_\Omega
}t\right) ^{\frac 23}
+\frac{K_1(m_1,x)}{M^{8/3}} \right] \label{eq:em}
\end{equation}
\begin{eqnarray}
K_1(m_1,x)&=&\left (m_1-1\right )F x {\it LerchPhi} (x,1,3m_1/5+1)-
\nonumber \\
& & \left (m_1-1\right )F{\it LerchPhi}(x,1,3m_1/5)
\end{eqnarray}
where,
\begin{equation}
F=\frac{2^{7/3} \pi^{2/3} \xi \rho_{\rm b}^{2/3}}{3^{2/3} H^2 \Omega}
\int_0^r \frac{{\cal L}^2 dr}{r^3}
\end{equation}
$m_1=5/(n+3)$, $t_\Omega =\frac{\pi \Omega _0}{H_o\left( 1-\Omega
_0-\Omega _\Lambda \right) ^{\frac 32}}$,
$x=1+(\frac{t_{\Omega}}{t})^{2/3}$ which is connected to mass by
$M=M_{\rm 0} x^{-3 m_1/5}$ (V2000) and $\xi=\frac{r_{ta}}{x_1}$,
where $r_{ta}$ is the turn-around radius and $x_1$ is defined by
the relation $M=\frac{4\pi\rho_bx^3_1}{3}$ with $\rho_b$ is the
background density. The {\it LerchPhi} function is defined as
follows:
\begin{equation}
LerchPhi(z,a,v)=\sum_{n=0}^{\infty} \frac{z^n}{(v+n)^a}
\end{equation}
The angular momentum, ${\cal L}$, acquired by the protostructure,
is calculated using  the same model (and same spectrum) as
described in Del Popolo \& Gambera (1998, 1999). More hints on the
model and some of the model limits can be found in Del Popolo,
Ercan \& Gambera (2001)). Then ,the virial theorem with the
surface pressure term correction, as in V2000, is used in order to
get a connection between the kinetic energy and temperature. We
utilise the usual relation:
\begin{equation}
\langle K \rangle= \frac{3 \tilde{\beta} M k T}{2 \mu m_{\rm p}}
\label {eq:conn}
\end{equation}
(AC),
where $k$ is the Boltzmann constant, $\mu=0.59$ is the mean molecular weight, $m_{\rm p}$ the proton mass and $\tilde{\beta}=\frac{\sigma_{\rm v}^2}{kT/\mu m_{\rm p}}$, being $\sigma_{\rm v}$ the mass-weighted mean
velocity dispersion of dark matter particles,
and $\tilde{\beta} = \beta[1+f(1/\beta-1)\Omega_{\rm b}/\Omega_{\rm m}]$,
%
%
where $f$ is the fraction of the baryonic matter in the hot gas, and $\Omega_{\rm b}$ is the density parameter of the
baryonic matter. In this way, we finally get:
\begin{equation}
kT=\frac {2}{5}a\frac{\mu m_p}{2\beta} \frac {m_1}{m_1-1}\left(
\frac{2\pi G}{t_\Omega }\right) ^{\frac{2}{3}}M^{\frac {2}{3}}
\left[ \frac {1}{m_1}+\left( \frac{t_\Omega }t\right) ^{\frac
{2}{3}} +\frac{K_1(m_1,x)}{M_0^{8/3}} \right] \label{eq:kT}
\end{equation}
where $a=\frac{\overline{\rho}}{2 \rho(R_{\rm vir})-\overline{\rho}}$ is the ratio between kinetic and total energy (V2000).
If $K_1=0$, Eq. (\ref{eq:em}) reduces to Eq. (10) of V2000. As stressed by V2000,
 some factors give rise to an higher value of $E/M$ with respect the case of the
  late-formation value. The $m_1/(m_1-1)$ value which accounts for the effect of early infall.
   The $1/m_1$ value in the square bracket of Eq. (\ref{eq:em}) which accounts for the cessation of cluster formation when $t>>t_{\rm \Omega}$. Finally in Eq. (\ref{eq:em}) a new term is present, which comes from the tidal interaction.

Using the relation $\Delta_{\rm vir}=\frac{8 \pi^2}{H t^2}$ (see V2000), and in the early-time limit: ($t<<t_{\Omega}$), Eq. (\ref{eq:kT}), reduces to:
\begin{equation}
kT=\frac {2}{5} \frac {m_1}{m_1-1}a\frac{\mu m_p}{2\beta}  G
M^{\frac {2}{3}} \left( \frac{4 \pi}{3} \rho_{\rm b} \Delta_{\rm
vir} \right)^{1/3} \label{eq:kT1}
\end{equation}
which, in the case $n \sim -2$, $a \sim 2$ is identical
to the late-formation formula, described in V2000 (see their Eq. (8)).
Normalising Eq. (\ref{eq:kT}) similarly to V2000, we get:
\begin{equation}
kT \simeq 8 keV \left(\frac{M^{\frac {2}{3}}}{10^{15}h^{-1}
M_{\odot}}\right) \frac{ \left[ \frac {1}{m_1}+\left(
\frac{t_\Omega }t\right) ^{\frac {2}{3}} +\frac{K_1(m_1,x)}{M^{8/3}}
\right] } { \left[ \frac {1}{m_1}+\left( \frac{t_\Omega
}{t_{0}}\right) ^{\frac {2}{3}}
 +\frac{K_0(m_1,x)}{M_0^{8/3}}
\right]
}
\label{eq:kT1}
\end{equation}
where $K_0(m_1,x)$ indicates that $K_1(m_1,x)$ must be calculated
assuming $t=t_0$

Eq. (\ref{eq:kT1}) when compared to the result of V2000 (Eq. 17) shows an additional term, mass dependent. This means that, as in the case of the top-hat model, the M-T relation is no longer self-similar showing a break at the low mass end (see next section).

\section{Appendix B: L-T relation in the MPEM}

The X-ray bolometric luminosity of a cluster is given by Eq. (\ref{eq:bol}), which in
CMT98 notation is:
\begin{equation}
L\propto \int_o^{r_2}\,n^2(r)\,T^{1/2}(r)\,d^3r~.
\end{equation}
Here $T(r)$ is temperature in the plasma and $r_2$
is the cluster boundary, that we take to be close to
the virial radius $R_{\rm vir} \propto M^{1/3}\,\rho^{-1/3}$, where
$\rho (z)\propto (1+z)^3$ is the DM density in the cluster,
proportional to the average cosmic DM density $\rho_u(z)$ at formation.
The infalling gas is  expected to become supersonic near $r_2$
(see, e.g., Perrenod 1980; Takizawa \& Mineshige 1998) so
that a shock front will form there. The conservations across the shock of mass,
energy and stresses yield the Rankine-Hugoniot conditions, i.e.,
the temperature and density jumps from the outer
values $T_1$ and $n_1$ to $T_2$ and $n_2$ just interior to $r_2$.
Then the luminosity may be rewritten in the form
\begin{equation}
L\propto r_2^3\;n_2^2\,T_2^{1/2}\;
\int_0^1 d^3x\,\Big[{n(x)\over n_2}\Big]^2\,
\Big[{T(x)\over T_2}\Big]^{1/2}~,
\end{equation}
 where $x\equiv r/r_2$.
$n_1$ is fixed by $n_1 \propto f_u \, \rho_u /m_p$,
in terms of the universal baryonic fraction $f_u$; whereas $T_1$ is determined  only  statistically,
through the diverse merging histories ending up in the mass $M$.
In sum, a given dark mass $M$ admits a set of ICP equilibrium states
characterized by different boundary conditions, each corresponding to a
different realization of the dynamical merging history.
It is the convolution over such set which provides the
average values of $L$ and $R_X$, and their scatter.
Following CMT98, the pre-shock temperature in a merging event is
that of the infalling gas, and if the latter is contained in a sufficient deep potential well,
$T_1$ is the virial
temperature $T_{1v}\propto \Delta m/r$
of the secondary merging partner; on using
$r\propto (\Delta m/\rho)^{1/3}$ this writes
\begin{equation}
k\,T_{1v}=4.5\,(\Delta m)^{2/3}\,(\rho/\rho_o)^{1/3}~{\rm keV},
\end{equation}
where the numerical
coefficient is taken from Hjorth, Oukbir \& van Kampen (1998),
and the masses $m=M/M_0$ are normalised to the current value $M_0=0.6 \times 10^{15} \Omega_0 h^{-1} M_{\odot}$ (i.e., the mass enclosed within a sphere of $8 h^{-1}$ Mpc),
so in the following the actual value of $T_1$ will be
\begin{equation}
T_1=max\,\big[T_{1v},T_{1*}\big] \footnote{An independent lower bound
$kT_{1*}\approx 0.5$ keV is provided by
preheating of diffuse external gas, due to
feedback energy inputs following star formation and evolution
all the way to supernovae (David et al. 1995;
Renzini 1997). }
\end{equation}
Given $T_1$, the boundary conditions
for the ICP in the cluster
is  set by the strength of the shocks separating the inner
from the infalling gas.
In the case of three degrees of freedom and
for a nearly hydrostatic post-shock condition with $v_2<< v_1$,
assuming the shock velocity to match the growth rate of the
virial radius $R_{\rm vir}(t)$:
\begin{equation}
kT_2={{\mu m_p v_1^2}\over 3}\Big[ {{(1+\sqrt{1+\epsilon})^2}\over 4}
+ {7\over{10}}\epsilon -{{3}\over {20}}{{\epsilon^2}\over{(1+\sqrt{1+
\epsilon})^2}}\Big]\, .
\label{eq:kkt}
\end{equation}
Cavaliere, Menci e Tozzi (1997) (CMT97).

Here $\epsilon\equiv 15 kT_1/4 \mu m_p v_1^2$ and $\mu$ is the average
molecular weight;  the inflow velocity $v_1$ is set
by the potential drop across the region of nearly free fall, to read
$v_1 \simeq \sqrt{-\phi_2/m_p}$ in terms of the potential $\phi_2$ at $r_2$.
In the case of strong shocks, appropriate to ``cold inflow", $\epsilon <<1$,
as in rich clusters accreting small clumps and diffuse gas, the approximation
\begin{equation}
k T_2 \simeq -\phi_2/3 +3 k T_1/2
\label{eq:kt}
\end{equation}
holds,
where $\phi_2$ is the gravitational potential energy at $r_2 \simeq R_{\rm vir}$.
For $\epsilon \geq 1$ the shock is weak, and $T_2 \simeq T_1$.
From $T_2$ and $T_1$,
the density jump at the boundary $n_2/n_1$ is found to read
(see CMT97)
\begin{equation}
{n_2\over n_1} =
2\,\Big(1-{T_1\over T_2}\Big)+\Big[4\,
\Big(1-{T_1\over T_2}\Big)^2 + {T_1\over T_2}\Big]^{1/2}~.
\end{equation}

Adopting the polytropic temperature description
$T(x)/T_2= [n(x)/n_2]^{\gamma-1}$, with the
index $\gamma$ in the range $1\leq \gamma\leq 5/3$, and that
the radius $r_2$ may be written in terms of temperature $T_{\rm v} \propto m/r_2$
and that $m \propto \rho r_2^3$, leading to $r_2 \propto (t/\rho)^{1/2}$,
the luminosity can be written in the form:

\begin{equation}
L\propto \Big({n_2\over n_1}\Big)^2\, m T_v^{1/2}\,\rho\,
\Bigg[{T_2\over T_v}\Bigg]^{1/2}\,
\overline{[n(r)/n_2\big]^{2+(\gamma-1)/2}} ~,
\label{eq:ll}
\end{equation}
where the bar denotes the integration
over the emitting volume $r^3\leq r_2^3$, and $\rho$ is the average
DM density in the cluster, proportional
to $\rho_u$ and so to $n_1$.

Eq. (\ref{eq:ll}) can be also cast in the form:
\begin{equation}
L\propto \Big({n_2\over n_1}\Big)^2\,
\rho\,
\Bigg[{T_2\over T_v}\Bigg]^{1/2}\,
\overline{[n(r)/n_2\big]^{2+(\gamma-1)/2}}
m^{4/3}
\sqrt {
\frac{
\frac{1}{m_{1}}+\left ({\frac {t_{{\Omega}}}{t}}
\right )^{2/3}+
\frac{K}{
\left ({\frac {m}{m_{{o}}}}\right )^{8/3}
}
}{
\frac{1}{m_1}+\left ({\frac {t_{{\Omega}}}{t}}\right )^{2/3}+{\frac
{K_{{0}}}{{m_{{0}}}^{8/3}}}}
}
\end{equation}
The ratio $n(x)/n_2$ is obtained from the hydrostatic equilibrium
$dP/m_p\,n\,dr=-G\,M(<r)/r^2=-d\phi/dr$ with the  polytropic
pressure $P(r)= kT_2\,n_2\,\big[{n(r)/n_2}\big]^{\gamma}$.  This
yields (see Cavaliere \& Fusco Femiano 1978; Sarazin 1988, and
bibliography therein) the profiles
\begin{equation}
{n(r)\over n_2}=\Big[{T(r)\over T_2}\Big]^{1/(\gamma-1)}=
\Big\{1+{\gamma-1\over \gamma}\,\beta\,
\big[\tilde{\phi}_2-\tilde{\phi}(r)\big]\Big\}^{1/(\gamma-1)}~,
\label{eq:dens}
\end{equation}
where $\tilde{\phi}\equiv \phi/\mu\,m_p\,\sigma_2^2$ is the potential
normalised  to the associated one-dimensional DM velocity dispersion at $r_2$.
The ICP disposition in eq. (11) relative to the DM depends on the parameter, already met previously:
\begin{equation}
\beta = \mu m_p \sigma_2/kT_2~,
\label{eq:mu}
\end{equation}
and is further modulated by the second parameter $\gamma$, to
yield as the latter increases flatter profiles $n(r)$ and steeper $T(r)$.
\footnote{For the King potential (see Sarazin 1988) and CMT97, with core radius $r_c=R_{v}/12$, $\beta(T)$ increases from $\beta \simeq 0.5$, for $T \simeq T_1$ to $\beta \simeq 0.9$ for $T>>T_1$. A similar result is obtained for a Navarro et al (1996) potential.
The other parameter $\gamma$ will be bounded according to CMT99,
The polytropic index $\gamma\geq 1$ describes the equation of state
for the ICP. An upper bound to it arises if
the overall thermal energy of the ICP is not to exceed its
gravitational energy.
The thermal and the gravitational energy are computed using the profiles
in Eq. (\ref{eq:dens}),   and their ratio is given in Fig. 4 of CMT99, to show that
the {\it upper} bound  $\gamma\leq 1.3$ holds.
It turns out that
observations by Markevitch et al. (1997) are consistent with the
$T(r)$  predicted when $\gamma = 1.2\pm 0.1$, in our allowed range.
Hereafter we shall focus on $\gamma=1.2$.}

The function $\beta(T)$ can be easily computed from Eq. (\ref{eq:kt}) for a given dark matter potential $\phi_2$
corresponding to $\rho(r)$. $\phi(r)$ and $\sigma(r)$ are obtained in agreement with
Navarro, Frenk \& White (1997).


\section{Appendix C: Calculation of the angular momentum}


The effect of tidal torques on structures evolution has been studied in
several papers especially in connection with the origin of galaxies
rotation (Hoyle 1949; Peebles 1969; White 1984; Ryden 1988 (hereafter R88);
Eisenstein \& Loeb 1995).

Following Eisenstein \& Loeb (1995), we separate the universe into two
disjoint parts: the collapsing region, characterised by having high density,
and the rest of the universe.
The boundary between these two regions is taken to be a
sphere centred on the origin.
As usual, in the following, we denote with $\rho({\bf x})$, being ${\bf x}$
the position vector, the density as function of space and
$\delta({\bf x})={\rho({\bf x})-\rho_{\rm b} \over \rho_{\rm b}}$.
The gravitational force exerted on the spherical central region by the external
universe can be calculated by expanding the potential, $\Phi({\bf x})$, in spherical harmonics.
Assuming that the sphere has radius $R$, we have:
\begin{equation}
\Phi ({\bf x})=\sum_{l=0}^\infty {4\pi  \over 2l+1}%
\sum_{m=-l}^l a_{\rm lm}(x)Y_{\rm lm}(\theta ,\phi )x^l 
\end{equation}
where $Y_{\rm lm}$ are spherical harmonics and the tidal moments,
$a_{\rm lm}$, are given by:
\begin{equation}
a_{\rm lm}(x)=\rho_{\rm b}\int_R^\infty Y_{\rm lm}(\theta ,\phi )\rho ({\bf s})
s^{-l-1}d^3s 
\end{equation}

In this approach the proto-structure
is divided into a series
of mass shells and the torque on each mass shell is computed separately. The
density profile of each proto-structure is approximated by the superposition
of a spherical profile, $\delta (r)$, and a random CDM
distribution, ${\bf %
\varepsilon (r)}$, which provides the quadrupole moment of
the proto-structure.
To the first order, the initial density can be represented by:
\begin{equation}
\rho ({\bf r})=\rho _{\rm b}\left[ 1+\delta (r)\right] \left[ 1+\varepsilon ({\bf
r})\right]  
\end{equation}
where
$ \varepsilon(\bf r)$
is given by:
\begin{equation}
\langle |\varepsilon _k|^2 \rangle = P(k) 
\end{equation}
being $ P(k)$ the power spectrum.
The torque on a thin spherical shell of internal radius $x$ is given by:
\begin{equation}
{\bf \tau}(x)=-{G M_{\rm sh} \over 4 \pi} \int \varepsilon({\bf x})
{\bf x} {\bf \times} {\bf \bigtriangledown} \Phi({\bf x}) d \Omega 
\label{eq:tauu}
\end{equation}
where $M_{sh}= 4 \pi \rho_{\rm b}\left[1+\delta(x)\right] x^2 \delta x$.
Before going on, I want to recall that we are interested in the
acquisition of angular momentum from the inner region, and
for this purpose we take account only
of the $l=2$ (quadrupole) term. In fact, the $l=0$ term produces no force, while the
dipole ($l=1$) cannot change the shape or induce any rotation of the
inner region. As shown by Eisenstein \& Loeb (1995), in the standard CDM
scenario the dipole is generated at large scales, so the object we are studying
and its neighbourhood move as bulk flow with the consequence that the
angular distribution of matter will be very small, then the dipole terms can be
ignored. Because of the isotropy of the random field, $\varepsilon(\bf x)$,
Equation~(\ref{eq:tauu}) can be written as:
\begin{equation}
<|{\bf \tau}|^2>=\sqrt(30) {4 \pi G \over 5}
\left[<a_{2m}(x)^2><q_{2m}(x)^2>-
<a_{2m}(x) q^{\ast}_{2m}(x)>^2
\right]^{1/2}   
\label{eq:tauuu}
\end{equation}
where $<>$ indicates a mean value of the physical quantity considered.
As stressed in the next section, following Eisenstein \& Loeb (1995),
the integration of the equations of motion shall
be ended at some time before the inner external tidal shell (i.e.,
the innermost shell of the part of the universe outside the sphere containing
the ellipsoid) collapses.
Then the inner region behaves as a density peak. This last
point is an important one in the development of the present paper.

An important question to ask, before going on, regards the role of
triaxiality of the ellipsoid (density peak)
in generating a quadrupole moment.
Equation~(\ref{eq:tauuu}) takes into account the quadrupole moment
coming from the secondary perturbation near the peak. The density
distribution around the inner region is characterised
by a mean spherical distribution, $\delta$, and
a random isotropic field. In reality the central region is a triaxial
ellipsoid. It is then important to evaluate the contribution to the quadrupole
moment due to the triaxiality.
Remembering that the quadrupole moments are given by:
\begin{equation}
q_{\rm 2m}= \int_{|{\bf r}|<R} Y_{2m}^{\ast}(\theta, \phi)
s^2 \rho({\bf s}) d^3 s= {x^2 M_{\rm sh} \over 4 \pi}
\int Y_{2m}^{\ast}(\theta, \phi) \varepsilon({\bf x}) d \Omega  
\label{eq:quad}
\end{equation}
and approximating the density profile as:
\begin{equation}
\delta({\bf x})= <\delta(x)>_{Spherical}+\nu f(x) A(e,p)  
\label{eq:deltt}
\end{equation}
being $<\delta(x)>_{Spherical}$ the mean spherical profile,
$\nu={\delta \over \sigma}$ the peak height and $\sigma$ the r.m.s.
value of $\delta$.
The function $A(e,p)$ of the triaxiality parameters, $e$ and $p$,
is given by:
\begin{equation}
A(e,p)=3 e(1-\sin^2{\theta}-\sin^2{\theta} \sin^2{\phi})+p(1-3 \sin^2{\theta} \cos^2{\phi}) 
\label{eq:aaa}
\end{equation}
while
the function $f(x)$ is given (R88) by:
\begin{equation}
f(x)={5 \over 2 \sigma} R^2_{\ast}
\left({1 \over x} {d \xi \over d x} -{1 \over 3}
\bigtriangledown^2 \xi \right) 
\end{equation}
where $\xi$, $\sigma$ and $R_{\ast}$ are respectively the two-point
correlation function, the mass variance and a parameter connected to the spectral moments
(see Bardeen et al. 1986, equation~(4.6d), hereafter BBKS). Substituting
equation~(\ref{eq:deltt}) and equation~(\ref{eq:aaa}) in
equation~(\ref{eq:quad}) it is easy to show that the sum of the
mean quadrupole moments due to triaxiality is:
\begin{equation}
{1 \over M_{\rm sh}} \sum_{m=-2}^{2} <q_{\rm 2m}(x)>= \nu x^2 f(x)
\left({1 \over 2 \pi} \sqrt{6 \pi/5} (e-p) +
{1 \over 4 \pi} \sqrt{4 \pi/5} (3e+p)
\right) 
\end{equation}
which must be compared with that produced by the secondary perturbations,
$\varepsilon$:
\begin{equation}
<q_{\rm 2m}(x)^2>={x^4 \over (2 \pi)^3} M^2_{\rm sh}
\int k^2 P(k) j_2(kx)^2 dk  
\end{equation}
where $j_2$ is the Bessel function of order 2.
The values of $e$ and $p$ can be obtained from
the distribution of ellipticity and prolateness (BBKS, equation~(7.6) and figure~7)
or for $\nu>2$
by:
\begin{equation}
e= {1 \over \sqrt{5} x \left[1+6/(5 x^2)\right]^{1/2}} 
\end{equation}
and
\begin{equation}
p= {6 \over 5 x^4 \left[1+6/(5 x^2)\right]^{2}} 
\end{equation}
(BBKS equation~(7.7)).
where $x$ is given in BBKS (equation~(6.13)).
In the case of a peak with $\nu=3$, we have $e \simeq 0.15$,
$p \simeq 0.014$ while for peaks having $\nu=2$ and $\nu=1$ they are
respectively given by $e \simeq 0.2$, $p \simeq 0.03$ and 
$e \simeq 0.25$ $p \simeq 0.04$.

As shown in figure~1 of Del Popolo et al. (2001), for a $3 \sigma$ profile,
the source of quadrupole moment due to triaxiality is less important than
that produced by the random perturbations $\varepsilon$ in all the
proto-structure, except
in the central regions
where the quadrupole moment due to triaxiality is comparable in magnitude
to that due to secondary perturbations.
In other words, the triaxiality has a significant effect only in the
very central regions, which contains no more than a few percent of the total
mass and where the acquisition of angular momentum is negligible. It follows
that the triaxiality can be ignored while computing both expansion and
spin growth (R88).
Moreover, as observed by Eisentein \& Loeb (1995), the ellipsoid model
does better in describing low shear regions (having higher values of $\nu$),
whose collapse is more spherical and then the effects of triaxiality are
less evident. Just this peaks, having at least $\nu>2$, shall be studied in this
paper.
In any case, even if the triaxiality was not negligible it should
contribute to increment the acquisition of angular momentum
(Eisenstein \& Loeb 1995), and finally to a larger effect on the density
evolution, (i.e., a larger reduction of the growing rate of the density).

In order to find the total angular momentum imparted to a mass shell by tidal
torques, it is necessary to know the time dependence of the torque.
This can be done connecting $q_{\rm 2m}$ and $a_{\rm 2m}$ to
parameters of the spherical collapse model (Eisenstein \& Loeb 1995
(equation~(32), R88 (equation~(32) and (34)). 
Following R88 we have:
\begin{equation}
q_{\rm 2m} (\theta )={1 \over 4} q_{\rm 2m,0}
\overline{\delta} _0^{-3} 
{\left( 1-\cos {\theta}\right) ^2 f_2 (\theta) \over
f_1(\theta )-\left({\delta _0 \over \overline{\delta} _0}\right)
f_2(\theta )}  
\label{eq:quadd}
\end{equation}
and
\begin{equation}
a_{\rm 2m} (\theta )=a_{\rm 2m,0} 
\left({4 \over 3}\right)^{4/3}
\overline{\delta}_0 (\theta-\sin{\theta})^{-4 \over 3} 
\label{eq:a2m}
\end{equation}
The collapse parameter $\theta$ is given
by:
\begin{equation}
t(\theta)={3 \over 4} t_0 \overline{\delta}_0^{-3/2}(\theta-\sin{\theta})
\end{equation}
Equation~(\ref{eq:quadd}) and (\ref{eq:a2m}),
by means of equation~(\ref{eq:tauuu}), give to us the tidal
torque:
\begin{equation}
\tau (\theta )=\tau _0 {1 \over 3}({4 \over 3})^{(1/3)}
\overline{\delta} _0^{-1}{
\left( 1-\cos {\theta }\right) ^2 \over (\theta -\sin {\theta })^{(4/3)}}
{f_2(\theta ) \over f_1(\theta )-
\left({\delta _0 \over \overline{\delta} _0}\right)
f_2(\theta )} 
\end{equation}
where $f_1(\theta)$ and $f_2(\theta)$ are given in R88 (Eq. 31), $\tau_0$
and $\delta_0={\rho-\rho_{\rm b} \over \rho_{\rm b}}$
are respectively the torque and the mean fractional density excess inside the shell,
as measured at current epoch $t_0$.
The angular momentum acquired during expansion can then be obtained integrating
the torque over time:
\begin{equation}
L=\int \tau(\theta) {d t \over d \theta} d\theta  
\label{eq:ang}
\end{equation}
As remarked in the Del Popolo et al. (2001) the angular momentum obtained from
equation(\ref{eq:ang}) is evaluated at the time of maximum expansion $t_{\rm M}$.
Then the calculation of the angular momentum can be solved by means
of equation~(\ref{eq:ang}), once we have made a choose for the power spectrum.
With the power spectrum and the parameters given in the next section and 
for a $\nu=2$ peak, the model gives a value of
$2.5 \times 10^{74} {\rm g cm^2/s}$. As previously quoted, we assume
that from $t_{M}$ on, the ellipsoid has this constant angular momentum. 
Following the procedures 1) and/or 2), we shall be able to get the
time evolution of the density.

---------------------------------------------------------------------

\newpage


\begin{thebibliography}{}
\bibitem{} Afshordi N., Cen R., 2001, astro-ph/0105020
\bibitem{} Allen S.W., Fabian A. 1998, MNRAS 297, L57
\bibitem{} Allen S.W., Schmidt R.W., Fabian A.C., 2001, MNRAS 328, 37
\bibitem{} Audit E., Teyssier R., Alimi J. M., 1997, A\&A 325, 439
\bibitem{} Balogh M.L., Babul A., Patton D.R., MNRAS 307, 463
\bibitem{} Bardeen J.M., Bond J.R., Kaiser N., Szalay A.S., 1986, ApJ 304, 15
\bibitem{} Barnes J., Efstathiou G., 1987, ApJ,319,575
\bibitem{} Barrow J.D., Silk J., 1981, ApJ 250, 432
\bibitem{} Bialek J.J., Evrard A.E., Mohr J.J., 2001, ApJ 555, 597, (astro-ph/0010584)
\bibitem{} Bond J.R., Myers S.T., 1993a, preprint CITA/{93}/27
\bibitem{} Bond J.R., Myers S.T., 1993b preprint CITA/{93}/28
\bibitem{} Borgani et al. 2001, ApJ 559, L71
\bibitem{} Bower R.G., Castander F.J., Couch W., Ellis R.S., B\"oringer H., 1997, MNRAS 291, 353
\bibitem{} Bryan G.L., Norman M.L., 1998, ApJ 495, 80, (astro-ph/9710107)
\bibitem{} Bryan G.L., ApJ 544, L1-L5
\bibitem{} Carlberg R.G., Yee H.K.C., Ellingson E., 1997, ApJ 478, 462
\bibitem{} Catelan P., Theuns T.,1996, MNRAS,282,436
\bibitem{} Cavaliere A., Menci N., Tozzi P., 1997, ApJ 484, 21
\bibitem{} Cavaliere A., Menci N., Tozzi P., 1998, ApJ 501, 493
\bibitem{} Cavaliere A., Menci N., Tozzi P., 1999, MNRAS 308, 599
\bibitem{} David L. P., Slyz A., Jones C., Forman W., Vrtilek S.D., 1993, ApJ, 412, 479
\bibitem{} David L.P., Jones C., Forman W., 1995, ApJ 445, 578
\bibitem{} Davis M., Peebles P.J.E., 1977, ApJS 34, 425
\bibitem{} Del Popolo A., Gambera M., 1998, A\&A 337, 96
\bibitem{} Del Popolo A., Gambera M., 1999, A\&A 344, 17
\bibitem{} Del Popolo, A., E. N. Ercan, Z. Q. Xia, 2001, AJ 122, 487
\bibitem{} Del Popolo A., 2002a A\&A 387, 759
\bibitem{} Del Popolo, A., 2002b, MNRAS 336, 81
\bibitem{} Edge A.C., Stuart G.C., 1991, MNRAS 252, 414
\bibitem{} Efstathiou G., Jones B.J.T., 1979, MNRAS,186,133
\bibitem{} Eisenstein D.J., Loeb A., 1995, ApJ,439,520
\bibitem{} Eke V.R., Cole S., Frenk C.S., 1996, MNRAS 282, 263
\bibitem{} Evrard A.E., Henry J.P., 1991, ApJ 383, 95
\bibitem{} Evrard A.E., Metzler C.A., Navarro J.F., 1996, ApJ 469, 494
\bibitem{} Finoguenov A., Reiprich T.H., B\"oeringer H., 2001, A\&A 368, 749
\bibitem{} Gioia I., Henry J.P., Maccacaro T., Morris S.L., Stocke J.T., Wolter A., 1990, ApJ 356, 35
\bibitem{} Henry J.P., Gioia I., Maccacaro T., Morris S.L., Stocke J.T., Wolter A., 1992, ApJ 386, 408
\bibitem{} Hernquist L., 1987
\bibitem{} Hjorth J., Oukbir J., van Kampen e., 1998, MNRAS 298, L1
\bibitem{} Hiotelis N., 2002, A\& A, 382,84
\bibitem{} Hoffman Y., 1986, ApJ 301, 65
\bibitem{} Horner D.J., Mushotzky R.F., Scharf C.A., 1999, Apj 520, 78
\bibitem{} Kaiser N., 1986, MNRAS 222, 323
\bibitem{} Kaiser N., 1991, ApJ 383, 104
\bibitem{} Kitayama T., Suto Y., 1996, ApJ 469,480
\bibitem{} Kratsov A.V., Klypin A.A., Bullock J.S., Primack J.R., 1998, ApJ, 502,48
\bibitem{} Lacey C., Cole S., 1993, MNRAS 262, 627
\bibitem{} Lokas E.L., Juskiewicz R., Bouchet F.R., Hivon E., 1996, ApJ 467, 1
\bibitem{} Markevitch M., 1998, Apj 503, 77
\bibitem{} Mathiesen B.F., Evrard A.E., 2001, astro-ph/0004309 (ApJ, in press)
\bibitem{} Mathiesen B.F., 2001, MNRAS 326, 1, (astro-ph/0012117)
\bibitem{} Muanwong O., Thomas P. A., Kay S. T., Pearce F. R., Couchman H. M. P., 2001, ApJ 552, 27, (astro-ph/0102048)
\bibitem{} Navarro J.F., Frenk C.S., White S.D.M., 1995, MNRAS 275, 720
\bibitem{} Navarro J.F., Frenk C.S., White S.D.M., 1997, ApJ 490, 493
\bibitem{} Neumann D.M., Arnaud M., 1999, A\&A 348, 711
\bibitem{} Nevalainen J., Markevitch M., Forman W., 2000, ApJ 532, 694
\bibitem{} Pearce F.R., Thomas P.A., Kay S.T., Pearce F.R., Couchman H.P.M., ApJ 552, L27-L30
\bibitem{} Peebles, P.J.E.,1971, A\& A,11,377
\bibitem{} Peebles P.J.E., Groth E.J., 1976, A\&A 53, 131
\bibitem{} Peebles P.J.E., 1990, ApJ 365, 27
\bibitem{} Peebles, P.J.E., 1993, Principles of Physical Cosmology, Princeton University Press
\bibitem{} Perrenod S.C. 1980, ApJ 236, 373
\bibitem{} Ponman T.J., Cannon D.B., Navarro J.F., 1999, Nature 397, 135
\bibitem{} Renzini A., 1997, ApJ 488, 35
\bibitem{} Ryden B.S., 1988, ApJ 329, 589
\bibitem{} Sarazin C.L., 1988, X-ray emission from Clusters of Galaxies (Cambridge: Cambridge University Press)
\bibitem{} Shimizu M., Kitayama T., Sasaki S., Suto Y., 2003, ApJ 590, 197
\bibitem{} Steinmetz M., Bartelmann M., 1995, MNRAS,272,570
\bibitem{} Szalay A.S., Silk J., 1983, ApJ 264, L31
\bibitem{} Takizawa M., Mineshige S., 1998, ApJ 499, 82
\bibitem{} Thomas P.A.,1998
\bibitem{} Thomas P.A., Muanwong O., Kay S.T., Liddle A.R.,2001, astro-ph/0112449
\bibitem{} Tozzi P., Norman C. 2001, ApJ 546, 63
\bibitem{} Viana P.T.P, Liddle A.R., 1996, MNRAS 281, 323
\bibitem{} Villumsen J.V., Davis M., 1986, ApJ 308, 499
\bibitem{} Voglis N., Hiotelis N., 1989, A\&A, 218,1
\bibitem{} Voit C. M., Donahue M., 1998, ApJ 500, 111 (astro-ph/9804306)
\bibitem{} Voit C. M., 2000, ApJ 543, 113 (astro-ph/0006366)
\bibitem{} Voit M., Bryan G. 2001, Nature 414, 425
\bibitem{} Warren M.S., Quinn P.J., Salmon J.K., Zurek W.H., 1992, ApJ,399,405
\bibitem{} White R.E., 1991, ApJ 367, 69
\bibitem{} Wu K.K.S., Fabian A.C., Nulsen P.E.J, 2000, MNRAS 308, 599
\bibitem{} Xu H., Jing G., Wu X., 2001, ApJ 553, 78, (astro-ph/0101564)
\end{thebibliography}
\end{document}